\documentclass[twocolumn,pre,showpacs,amsmath]{revtex4}
\usepackage{epsfig,amsfonts,amsbsy,color}
\newcommand{\bra}{\left\langle}
\newcommand{\ket}{\right\rangle}

\newcommand{\state}[1]{\left \vert \left. #1 \right\rangle   \right.}
\newcommand{\bstate}[1]{\left\langle  \left.  #1  \right \vert \right. }
\newcommand{\braket}[2]{\left\langle   #1 \vert #2 \right\rangle }

\newcommand{\nstate}[2]{\left \vert \left. #1 \right\rangle_{#2}   \right.}

\newcommand{\bnbraket}[3]{\hspace{-5pt}{\phantom{\rangle}}
                  \hspace{-2pt}\left\langle   #1 \vert #2 \right\rangle_{#3} }

\newcommand{\pder}[2]{\frac{\partial #1}{\partial  #2}}
\newcommand{\pderf}[3]{\left(\frac{\partial #1}{\partial  #2}\right)_{#3}}

\newcommand{\pdertf}[3]{\left( \frac{\partial^2 #1}{\partial  #2^2}\right)_{#3}}
\newcommand{\pderc}[3]{\frac{\partial^2 #1}{\partial  #2 \partial  #3}}

\newcommand{\pdert}[2]{\frac{\partial^2 #1}{\partial  #2^2}}
\newcommand{\der}[2]{\frac{d #1}{d  #2}}

\newcommand{\bv}[1]{{\boldsymbol #1}}

\newcommand{\nm}{\nonumber\\}

\begin{document}
\title{Thermodynamical path integral and emergent symmetry}

\author{Shin-ichi Sasa}
\affiliation {
Department of Physics, Kyoto University, Kyoto 606-8502, Japan}

\author{Sho Sugiura}
\affiliation {
Department of Physics, Harvard University, Cambridge, MA 02138, USA
}

\author{Yuki Yokokura}
\affiliation{iTHEMS Program, RIKEN, Wako, Saitama 351-0198, Japan}

\date{\today}

\begin{abstract}
  We investigate a thermally isolated quantum many-body system
  with an  external control represented by a step protocol of a parameter.
  The propagator at each step of  the parameter change
  is described by thermodynamic quantities under some assumptions.
  For the time evolution of such systems, we formulate a path integral over the trajectories in the thermodynamic state space.
  In particular, for quasi-static operations, 
  we derive an effective action of the thermodynamic entropy and its canonically conjugate variable. 
  Then, the symmetry for the uniform translation of the conjugate variable emerges in the path integral. 
  This leads to the entropy as a Noether invariant in quantum mechanics. 
\end{abstract}


\maketitle

\section{Introduction}

Thermodynamics and quantum mechanics are fundamental
theories in physics. The universal behavior of macroscopic
objects is described
by thermodynamics, while the microscopic dynamics of any system is governed
ultimately by quantum mechanics. Statistical mechanics connects
them in equilibrium states; 
however, the relation between their 
dynamics is not established 
despite studies in many contexts, such as
thermodynamic processes in quantum systems
\cite{Esposito,Campisi, Oppenheim,Wehner, Ikeda, IKS} 
and relaxation of pure quantum states to the thermal equilibrium
\cite{Neumann, Rigol,Reimann09,  Winter, Biroli, PolkovnikovReview2011,
GHT, GHT2,Eisert_Review2015}. 
Recently, state-of-the-art experiments for these studies are realized
by using ultracold atoms  \cite{exp-1,exp-2,Gring1318,MeasurementEE2015},
nuclear magnetic resonance \cite{Batalhao}, trapped ions \cite{Am2014},
and electronic circuits \cite{Pekola2015}.
Given these backgrounds, we propose a theory connecting thermodynamical
behavior to quantum mechanics.

Our strategy is to construct a thermodynamical path integral. 
In thermodynamics, an equilibrium state of a system is represented
by a point in the thermodynamic state space. In quantum mechanics,
on the other hand, the time evolution of a system is formulated in terms
of a sum over all possible paths in a configuration space,
weighted by the exponent of the action. In this paper, we combine
these two concepts for a thermally isolated quantum many-body
system under a time-dependent external control. We formulate the unitary
evolution of quantum states by an integral over paths in the thermodynamic
state space. 


The path integral is constructed as follows.
First, we introduce the projection operator
to an energy shell that constitutes
the micro-canonical ensemble of any energy $E$. 
In terms of the projection operators 
we express a decomposition of the identity operator. 
Next, we consider a series of step operations with the external control. 
At each step, the change of energy is much smaller than the energy itself but larger than the fluctuation. 
Then, we insert the decomposition formula at each step, 
evaluate a one-step propagator,
and take a continuum limit.

We show that the propagator is expressed
in terms of thermodynamic quantities.
This is the key of our derivation. 
In order to evaluate the propagator, we
introduce two assumptions. 
The first assumption is that
the time interval between two successive step operations
is so long that the phase of each energy eigenstate in the evolved state
is interpreted as a uniform random variable.
Now, for the time-evolved state, the state after the projection to an energy
shell has the amplitudes of each energy eigenstate in
the shell. The second assumption is that 
the amplitudes are equally weighted at each step. 
Such a class of non-equilibrium processes can be described 
by the thermodynamical path integral, 
which connects the concepts of thermodynamics
and quantum mechanics in dynamical problems. 

For quasi-static operations, 
we derive an effective action 
that has a symplectic structure for $(S,\hbar \theta)$, 
where $S$ is the thermodynamic entropy and $\theta$ is an auxiliary variable introduced in the path integral. 
The equations of motion are $dS/dt=0$ and $d \theta/dt=1/(\hbar \beta)$,
where $\beta$ is the inverse temperature depending on time. 
In such slow operations, 
the symmetry for $\theta \to \theta +\eta$ emerges in the path integral, 
leading to entropy conservation in quantum mechanics, 
where $\eta$ is an infinitely small parameter. 
This provides a complementary view to the quantum adiabatic theorem
\cite{Tolman,Kato} because the operations are assumed to be slow yet so
fast that transitions between different energy levels occur. 


This emergent symmetry is related to the following topics.
First, $\theta$ corresponds to a {\it thermal time}, which
was introduced as a parameter of the flow determined
by a statistical state 
\cite{Rovelli93,Connes-Rovelli94,Roveili-Smerlak,Haggard}. 
Through the relation  $dt= \hbar \beta d\theta$, 
the symmetry of the effective action for $\theta \to \theta + \eta$ 
is connected to that for $t \to t +\eta \hbar \beta$, which leads to 
entropy in classical systems \cite{Sasa-Yokokura}. 
Second, a similar symmetry has been
phenomenologically studied for perfect fluids \cite{Brown93,Kambe}
and for effective field theories \cite{adiabatic-hydrodynamics,deBoer,Liu}.
Finally, the entropy of stationary black holes is derived as the Noether
charge for $v \to v +\eta \hbar \beta_H$, where $v$ is the Killing parameter
and $1/\beta_H$ is the Hawking temperature \cite{Wald}.
Thus, our theory provides a unified perspective for studying the thermal time, 
perfect fluids, and black holes in terms of quantum mechanics. 


This paper is organized as follows.
In Sec.~\ref{setup}, we start with our setup. 
We describe isolated quantum many-body systems and 
employ an energy eigenstate as the initial state. 
In Sec.~\ref{TPSP}, we introduce 
a decomposition of the identity operator using projection operators onto energy shells,  
which plays a key role in our theory. 
By using this and the two assumptions, 
we formulate the thermodynamical path integral. 
In Sec.~\ref{TEA}, 
we construct quasi-static operations 
and derive the effective action in the thermodynamical phase space $(S,\hbar \theta)$. 
Then, we show the symmetry for $\theta \rightarrow \theta+\eta$
and the entropy conservation, and discuss characterizations of
the variable $\theta$. 
In the final section, we provide concluding remarks. 

\section{Setup}\label{setup}
Although the theory developed in this paper is applicable to a wide
class of quantum many-body systems, we specifically consider a
Hamiltonian $\hat H( h )$ consisting of $N$ spins with spin-1/2
under a uniform magnetic field $h>0$ so that the argument is explicit. 
We also assume that the system 
does not have any conserved quantities for any value of $h$.
It is straightforward to extend our result to the case 
with the existence of a small number of conserved quantities 
such as momentum and particle number \cite{conserved quantities}.
The eigenvalues and eigenstates satisfy
\begin{equation}
  \hat H(h) \state{n,h} =E(n,h) \state{n,h},
\end{equation}  
where $n=1, 2, \cdots, 2^N$. 
By incorporating the magnetic moment into $h$, 
we assume the dimension of  $h$ to be energy.   
Then, $h$ represents 
the characteristic energy scale per
unit spin. We study the macroscopic behavior of the
system by taking the large $N$ limit. 

We choose an energy shell $I_E\equiv [E-\Delta/2, E+\Delta/2]$,
where  $\Delta$ is much smaller than $hN$ but should be
large so that $I_{E}$ contains $e^{O(N)}$ energy levels.
Here we choose $\Delta=O(1)$ as an example.
The number of eigenvalues in the shell is given by 
$\sum_n \chi_{I_{E}}(E(n,h)) $, where
$\chi_{I_E}(x)=1$ for $x \in I_E$ and zero otherwise. 
The density
of states, $D(E,h)$, is defined as 
\begin{equation}
  D(E,h)\equiv \sum_n \frac{1}{\Delta}\chi_{I_E}(E(n,h)).
\end{equation}  
We assume the asymptotic form for large $N$
\begin{equation}\label{D}
  D(E,h)= e^{ N s(E/N, h)+o(N)}
  \end{equation}
with a function $s(u,h)$ whose functional form
is independent of $N$. 
This assumption is necessary for the consistency
of statistical mechanics with thermodynamics.
In fact, (\ref{D}) is satisfied for a wide class of systems
with local interactions. For thermodynamic states $(E,h)$, the entropy $S(E,h)$ is then defined as
\begin{equation}
  S(E,h)\equiv Ns(E/N, h).
\end{equation}  
The inverse temperature $\beta(E,h)$ is 
defined by the thermodynamic relation
\begin{equation}
  \beta\equiv \left( \pder{S}{E} \right)_h.
\end{equation}  
The Boltzmann constant is set to unity.

We consider a time-dependent magnetic field $h(t)$ in $0 \le t \le t_f$.
In particular, we employ a 
step protocol $h(t)=h_j$ for
$t_j \le t \le t_{j+1}$, where $t_j=j \Delta t$ and $t_f=M \Delta t$.
We choose $h_j$ such that $\Delta h_j\equiv h_{j}-h_{j-1}$ satisfies
\begin{equation}
  \frac{1}{\sqrt{N}} \ll \frac{|\Delta h_j|}{h_j}  \ll 1.
\label{m-per}
\end{equation}  
This means that the change of energy caused by the parameter change, which is $O(N\Delta h_j)$,  
is much smaller than the energy itself, but it is 
larger than the fluctuation for large $N$. 
In this sense, each quench is called {\it small macroscopic}. 
It should be noted here that the standard perturbation
technique cannot be employed for this protocol. 
Under this external field, 
the time evolution of a given initial state
$\state{\Psi(0)}$ is determined by 
\begin{equation}\label{Sch_eq}
	i \hbar \der{}{t}\state{\Psi(t)}=\hat H(h(t))\state{\Psi(t)}. 
\end{equation}

We study cases where 
the system is in a thermal equilibrium state at $t=0$.
We express the state by a single pure state, as per previous studies 
\cite{Neumann,Lloyd,Tasaki, GLTZ,Popescu,SugitaE, Reimann,Sugiura-Shimizu,Sugiura-Shimizu-2}. 
Unitary time evolution starting from such a thermal pure state 
is determined by (\ref{Sch_eq}), which is in accordance with isolated
quantum systems \cite{exp-1,exp-2} and may provide an idealization
of quantum dynamics in nature. 
In particular, we set
\begin{equation}
  \state{\Psi(0)}= \state{n_0,h_0}.
\label{in_state}
\end{equation}
Here, it should be noted that any single energy eigenstate
may exhibit a thermal equilibrium state, according
to the eigenstate thermalization hypothesis \cite{Rigol}.

\section{Thermodynamical path integral}
\label{TPSP}
In this section, we express the time-evolved state $\state{\Psi(t)}$ as 
a path integral over trajectories in the thermodynamic state space $(E,h)$. 
First, we use a formula of decomposition of identity operator $\hat 1$ and
construct a path-integral-like form of $\state{\Psi(t)}$. 
Then, we introduce two assumptions, which enable us to evalurate the propagator, 
and we express it in terms of thermodynamic quantities. 
Finally, we reach the path-integral expression. 

\subsection{Projected states} 
We first express the identity operator $\hat 1$ 
as an integration over energy $E$. 
We define a projection operator to $I_E$
\begin{equation}
\hat{\mathcal{P}}_{E,h}\equiv
\sum_n \chi_{I_{E}} (E(n,h))\state{n, h} \bstate{n, h},
\label{Projection}
\end{equation}
and we note that 
\begin{equation}
1= \frac{1}{\Delta}\int_{{\cal D}(h)} dE \chi_{I_E}(E(n,h))
\label{unity}
\end{equation}
holds for each $n$. 
Here the interval of integration is defined as 
\begin{align}
{\cal D}(h) \equiv [E_{\min}(h) - \Delta/2,E_{\max}(h) + \Delta/2], 
\end{align}
where $E_{\max}(h)$ and $E_{\min}(h)$ are 
the maximum and minimum energy eigenvalues for a given $h$, respectively. 
Using (\ref{unity}), 
we can express the complete relation $\hat 1=\sum_n\state{n, h} \bstate{n, h}$ as 
\begin{equation}
\hat{1}=
	{1\over \Delta}
	\int_{{\cal D}(h)} 
	\hat{\mathcal{P}}_{E,h} dE
\label{id_P}
\end{equation}
for each $h$.

Let us start with the evolution of (\ref{Sch_eq}) for $t_f=2\Delta t$:
\begin{align}
 \state{\Psi(2\Delta t)} &= 
  e^{-\frac{i}{\hbar} \hat H_1 \Delta t} e^{-\frac{i}{\hbar}\hat H_0 \Delta t} \state{n_0,h_{0}} \nonumber\\
 &= e^{-\frac{i}{\hbar} \hat H_1 \Delta t} e^{-\frac{i}{\hbar}E(n_0,h_0)\Delta t} \state{n_0,h_{0}}\nonumber\\
 &= e^{-\frac{i}{\hbar} (E_0 +\hat H_1) \Delta t}\int_{{\cal D}(h_1)}\frac{dE_1}{\Delta}\hat{\mathcal{P}}_{E_1,h_1}\state{n_0,h_{0}}, 
\label{path-integral 3}
\end{align}
where we have set $\hat H_j=\hat H(h_j)$ and $E_0=E(n_0,h_0)$
and used (\ref{id_P}) for $h_1$. 
Here, we define a projected state
\begin{equation}
\nstate{\mathcal{P}_{\bv{E}_1,\bv{h}_1}}{\Delta t}\equiv
\frac{\hat{\mathcal{P}}_{{E}_1,h_1}\state{n_0,h_0}}
{\sqrt{B_1({\bv{E}}_1,{\bv{h}}_1)}} 
\end{equation}
with the notation ${\bv{E}}_1\equiv(E_0,E_1), {\bv{h}}_1\equiv(h_0,h_1)$ and 
the normalization factor
\begin{equation}
B_1({\bv{E}}_1,{\bv{h}}_1) \equiv \bstate{n_0,h_0}
  \hat{\mathcal{P}}_{E_1,h_1}\state{n_0,h_0}.
\end{equation}
Then, (\ref{path-integral 3}) is expressed as 
\begin{align}
 \state{\Psi(2\Delta t)}  &= e^{-\frac{i}{\hbar} (E_0+\hat H_1) \Delta t}\nonumber \\
 &~~~\int_{{\cal D}(h_1)}  \frac{dE_1}{\Delta}\sqrt{B_1({\bv{E}}_1,{\bv{h}}_1)}\nstate{\mathcal{P}_{{\bv{E}}_1,{\bv{h}}_1}}{\Delta t}.
\label{path-integral 4}
\end{align}
Now, we suppose that 
$\nstate{\mathcal{P}_{{\bv{E}}_1,\bv{h}_1}}{\Delta t}$
evolves by an
energy-shifted Hamiltonian $\hat H_1-E_1$ during $[\Delta t,2\Delta t]$ to
\begin{align}
&\nstate{\mathcal{P}_{{E}_1,h_1}}{2\Delta t} 
\equiv e^{-\frac{i}{\hbar}(\hat{H}_1-E_1)\Delta t}
\nstate{\mathcal{P}_{{E}_1,h_1}}{\Delta t}.
\end{align}
Thus, we reach  
\begin{align}
\state{\Psi(2\Delta t)}
&=\int_{{\cal D}(h_1)} \frac{dE_1}{\Delta}\sqrt{B_1({\bv{E}}_1,{\bv{h}}_1)}\nonumber \\
 &~~~~~~~~~~\nstate{\mathcal{P}_{{\bv{E}}_1,{\bv{h}}_1}}{2\Delta t}
 e^{-\frac{i}{\hbar}(E_1+E_0)\Delta t}.
\label{path-integral 5}
\end{align}

By repeating this procedure, 
we can construct the general form for any $M$. 
To do it, we use the notation 
\begin{equation}
{\bv{E}}_{j}  \equiv  (E_0,\cdots, E_j ),~~{\bv{h}}_{j}  \equiv  ({h}_0,\cdots, {h}_j ),
\end{equation}
and define the projected state 
$\nstate{\mathcal{P}_{{\bv{E}}_j,{\bv{h}}_j}}{{j}\Delta t}$ in the
following iterative manner:
\begin{equation}
\nstate{\mathcal{P}_{{\bv{E}}_0,{\bv{h}}_0}}{0}
\equiv \state{n_0,h_0},
\end{equation}
\begin{equation}
\left|\mathcal{P}_{{\bv{E}}_j,{\bv{h}}_j} \right\rangle_{j\Delta t}
\equiv {1\over \sqrt{B_j({\bv{E}}_j,{\bv{h}}_j) }}\hat{\mathcal{P}}_{E_j,h_j}
\nstate{\mathcal{P}_{{\bv{E}}_{j-1},{\bv{h}}_{j-1}}}{j\Delta t},
\label{statePj}
\end{equation}
with 
\begin{equation}
B_j({\bv{E}}_j,{\bv{h}}_j) \equiv
_{j \Delta t}\bstate{\mathcal{P}_{{\bv{E}}_{j-1},{\bv{h}}_{j-1}}}
{\hat{\mathcal{P}}_{E_j, h_j}}
	\state{\mathcal{P}_{{\bv{E}}_{j-1},{\bv{h}}_{j-1}}}_{j\Delta t},
\label{Bjdef}
\end{equation}
and
\begin{eqnarray}
&\nstate{\mathcal{P}_{{\bv{E}}_j,{\bv{h}}_j}}{{(j+1)}\Delta t}
\equiv e^{-{i \over \hbar} (\hat{H}_j - E_j)
  \Delta t}\nstate{\mathcal{P}_{{\bv{E}}_j,{\bv{h}}_j}}{{j}\Delta t}.
\label{Pj+1}                
\end{eqnarray}
Then, we can reexpress (\ref{path-integral 5}) as
\begin{align}
\state{\Psi(2\Delta t)}
&=\int_{{\cal D}(h_2)} \frac{dE_2}{\Delta}\hat{\mathcal{P}}_{E_2,h_2}
\int_{{\cal D}(h_1)} \frac{dE_1}{\Delta}\nonumber \\
 &~~~~~\sqrt{B_1({\bv{E}}_1,{\bv{h}}_1)}\nstate{\mathcal{P}_{{\bv{E}}_1,{\bv{h}}_1}}{2\Delta t}
 e^{-\frac{i}{\hbar}(E_1+E_0)\Delta t}\nonumber \\
&=\int_{{\cal D}(h_2)} \frac{dE_2}{\Delta}
\int_{{\cal D}(h_1)} \frac{dE_1}{\Delta}e^{-\frac{i}{\hbar}(E_1+E_0)\Delta t}\nonumber \\
 &\sqrt{B_1({\bv{E}}_1,{\bv{h}}_1)}\sqrt{B_2({\bv{E}}_2,{\bv{h}}_2)}
 \nstate{\mathcal{P}_{{\bv{E}}_2,{\bv{h}}_2}}{2\Delta t},
\end{align}
where in the first line we insert (\ref{id_P}) for $h_2$ and in the second line use (\ref{statePj}). 
This is the path-integral-like representation of $\state{\Psi(2\Delta t)}$. 
Now, by repeating the above procedure, we can obtain the formula for $t_f=M\Delta t$: 
\begin{align}
\state{\Psi(t_f)}
=&
\left[ \prod_{j=1}^{M} \int_{{\cal D}(h_j)}
  \frac{d E_j}{\Delta}
\sqrt{B_j({\bv{E}}_j,{\bv{h}}_j)}
\right] \nonumber \\
&e^{-\frac{i}{\hbar} \sum_{j=0}^{M-1}E_j\Delta t}\nstate{\mathcal{P}_{{\bv{E}}_M,{\bv{h}}_M} }{M\Delta t}.
\label{mc path bf}
\end{align}

Here, $\sqrt{B_j({\bv{E}}_j,{\bv{h}}_j)}$ in (\ref{mc path bf}) can be expressed,
from (\ref{statePj}), (\ref{Bjdef}) and (\ref{Pj+1}), as
\begin{align}
&\sqrt{B_j({\bv{E}}_j,{\bv{h}}_j)} \nonumber\\
 =&\left|_{j \Delta t}\big\langle\mathcal{P}_{{\bv{E}}_{j},{\bv{h}}_{j}} \state{\mathcal{P}_{{\bv{E}}_{j-1},{\bv{h}}_{j-1}}}_{j\Delta t}\right|\nonumber\\
 =&\left|_{j \Delta t}\bstate{\mathcal{P}_{{\bv{E}}_{j},{\bv{h}}_{j}}}
{e^{-\frac{i}{\hbar}(\hat H_{j-1}-E_{j-1})\Delta t}}
	\state{\mathcal{P}_{{\bv{E}}_{j-1},{\bv{h}}_{j-1}}}_{(j-1)\Delta t}\right|.
\label{propa}
\end{align}
This is the propagator from $\state{\mathcal{P}_{{\bv{E}}_{j-1},{\bv{h}}_{j-1}}}_{(j-1)\Delta t}$
to $\state{\mathcal{P}_{{\bv{E}}_{j},{\bv{h}}_{j}}}_{j\Delta t}$, 
and (\ref{mc path bf}) looks like the standerd form of a path integral over $(E_1, E_2, \cdots, E_{M})$.
Note that, by the above construction, this propagator depends on the trajectory of
thermodynamic states $(E_0,h_0),\cdots, (E_{j-2},h_{j-2})$ in
addition to the thermodynamic states $(E_{j-1},h_{j-1}), (E_{j},h_{j})$. 

\subsection{Evaluation of $B_j$}\label{evaluation b}

We here show that the path dependence of 
$\sqrt{B_j({\bv{E}}_j,{\bv{h}}_j)}$ becomes negligible under two assumptions; 
we express $\sqrt{B_j({\bv{E}}_j,{\bv{h}}_j)}$ only in terms of $E_{j-1}, E_j ,h_{j-1},h_j$. 
To be specific, we represent $B_{j+1}$ in terms of $q_{nm}$ defined by 
\begin{equation}
q_{nm} \equiv \chi_{I_{E_{j+1}}}(E(n,h_{j+1}))
		\chi_{I_{E_j}}(E(m,h_j))
		\braket{n,h_{j+1}}{m,h_{j}}.
\label{qdef new}
\end{equation}                
Although $q_{nm}$ depends on $(E_j, E_{j+1},
h_j, h_{j+1})$, we do not write this dependence explicitly.
We then expand
$\nstate{\mathcal{P}_{{\bv{E}}_j,{\bv{h}}_j}}{j\Delta t}$ as 
\begin{equation}
\nstate{\mathcal{P}_{{\bv{E}}_j,{\bv{h}}_j}}{j\Delta t}
= \sum_n\chi_{I_{E_j}}(E(n,h_j)) d_{nj}({\bv{E}}_{j}, {\bv{h}}_{j})\state{n,h_j}.
\label{def_d}
\end{equation}
By substituting this into (\ref{Pj+1}), we have
\begin{align}
	\nstate{\mathcal{P}_{{\bv{E}}_j,{\bv{h}}_j}}{{(j+1)}\Delta t}
	= \sum_n &\chi_{I_{E_j}}(E(n,h_j))
        e^{-{i \over \hbar} (E(n,h_j) - E_j) \Delta t} \nm 
	& d_{nj}({\bv{E}}_{j}, {\bv{h}}_{j})\state{n,h_j}.
\label{p-exp}
\end{align}
Combining this expression with 
(\ref{Bjdef}) leads to 
\begin{align}
	B_{j+1}({\bv{E}}_{j+1},{\bv{h}}_{j+1})
	=& \sum_{n} \chi_{I_{E_{j+1}}}(E(n,h_{j+1}))
		 \nm
	&	\Big| \sum_m \chi_{I_{E_j}}(E(m,h_j))e^{-{i \over \hbar} (E(m,h_j) - E_j) \Delta t} \nm
	& \braket{n,h_{j+1}}{m,h_{j}}d_{mj}({\bv{E}}_{j}, {\bv{h}}_{j}) 
		\Big|^2 \nonumber \\
	        =& \sum_{n}
                \left|\sum_m e^{i \xi_{mj}} q_{nm} d_{mj}({\bv{E}}_{j}, {\bv{h}}_{j}) \right|^2\nonumber\\ 
        =& \sum_{n} |C_{nj}|^2,
\label{B-exp}                
\end{align}
where we have defined 
\begin{equation}
\xi_{mj} \equiv -{1 \over \hbar} (E(m,h_j) - E_j) \Delta t,
\label{xi-def}
\end{equation}
and 
\begin{align}
  C_{nj} \equiv \sum_m e^{i \xi_{mj}} q_{nm} d_{mj}({\bv{E}}_{j}, {\bv{h}}_{j}).
\label{C-exp}
\end{align}

We first study the properties of $\xi_{nj}$, 
which is order of $(\Delta t) \Delta/\hbar$. 
In (\ref{p-exp}) for a fixed $j$, a series of $E(n,h_j) (\in I_{E_j})$ in terms of $n$ is irregular 
as if it would follow some probability distribution. 
Therefore, when we choose a large $\Delta t$ satisfying 
\begin{equation}
\Delta t\gg \frac{\hbar}{\Delta},
\label{large_dt}
\end{equation}
we can assume that $\xi_{nj}$ are independent uniform random variables on $[0,2\pi]$. 
We then take the expectation value of $\left|C_{nj} \right|^2$ 
with respect to the random variables $\xi_{mj}$ $(m=1,\cdots,2^N)$ as follows \cite{comment_d}. 
\begin{align}
 \overline{ \left|C_{nj} \right|^2 } =&  
 \overline{\left| \sum_m e^{i \xi_{mj}} q_{nm}d_{mj} \right|^2 }  \nonumber \\
=&\sum_{m, m'}\overline{ e^{i (\xi_{mj}-\xi_{m'j}) }} q_{nm}d_{mj}q^*_{nm'}d^*_{m'j}  \nonumber \\
=& \sum_{m}  \left| q_{nm}d_{mj}  \right|^2.
\label{ex-rs-exp}
\end{align}
We can also show that 
\begin{align}
	\log |C_{nj}| &=\log \overline{|C_{nj}|}+o(N).
	\label{log B}
\end{align}
The precise statement is stated as (\ref{LL}), 
and the proof is given in Appendix \ref{PSA}.
Therefore, (\ref{B-exp}) becomes 
\begin{align}
B_{j+1}({\bv{E}}_{j+1},{\bv{h}}_{j+1})&=\sum_n  \overline{ \left|C_{nj} \right|^2 } e^{o(N)} \nonumber\\
 &=\sum_{n,m} \left| q_{nm} \right|^2 \left|d_{mj}({\bv{E}}_{j}, {\bv{h}}_{j}) \right|^2e^{o(N)}.
\label{B_exp2}
\end{align}

Next, we consider the form of $\left|d_{mj}({\bv{E}}_{j}, {\bv{h}}_{j}) \right|$.
From (\ref{statePj}), (\ref{propa}) and (\ref{def_d}), we can obtain 
\begin{equation}
\left|d_{mj}({\bv{E}}_{j}, {\bv{h}}_{j}) \right|^2
= \frac{\left| \langle {m,h_j}
	\state{\mathcal{P}_{{\bv{E}}_{j-1},{\bv{h}}_{j-1}}}_{j\Delta t}\right|^2 }
        {\left|_{j \Delta t}\big\langle\mathcal{P}_{{\bv{E}}_{j},{\bv{h}}_{j}}
        \state{\mathcal{P}_{{\bv{E}}_{j-1},{\bv{h}}_{j-1}}}_{j\Delta t}\right|^2}.
\label{d_exp}
\end{equation}
This is the probability of transition
from 
$\nstate{\mathcal{P}_{{\bv{E}}_{j-1},{\bv{h}}_{j-1}}}{{j}\Delta t}$ 
to $|m,h_j\rangle$ having $E(m,h_j)\in I_{E_j}$
when the quench $h_{j-1}\to h_j$ is performed at $t=j\Delta t$.  
The width of an energy shell is $\Delta=O(1)$,
while the quench is macroscopic in the sense that $N|h_j-h_{j-1}|\gg O(\sqrt{N})$ because of (\ref{m-per}).
Therefore, the transition can occur to any $|m,h_j\rangle$ in $I_{E_j}$ \cite{Santos}. 
In addition, the energy shell contains $e^{O(N)}$ energy eigenstates 
which do no have any particular structure. 
Motivated by this observation, we assume  
\begin{align}
	\log {|d_{mj}({\bv{E}}_{j}, {\bv{h}}_{j})|^2}
	=-\log D(E_{j}, h_{j})+o(N),
	\label{ETH}
\end{align}
which means that the overlap between 
$\nstate{\mathcal{P}_{{\bv{E}}_j,{\bv{h}}_j}}{j\Delta t}$
and each eigenstate in the shell is equally weighted
up to the sub-exponential factor in $N$. 
Thus, (\ref{B_exp2}) becomes 
\begin{equation}
B_{j+1}({\bv{E}}_{j+1},{\bv{h}}_{j+1})
=\frac{\sum_{n,m} \left| q_{nm} \right|^2}{D(E_j,h_j)} e^{o(N)}, 
\label{B_exp3}
\end{equation}
which depends only on $E_j,E_{j+1},h_j,h_{j+1}$.

\subsection{Expression of $\sum_{n,m}|q_{nm}|^2$ }
In this subsection, we evaluate $\sum_{n,m}|q_{nm}|^2$. 
Here we set $E=E_j$, $E'=E_{j+1}$, $h=h_j$, and $h'=h_{j+1}$. 
The key idea is that 
we express $\sum_{m,n} |q_{mn} |^2$ 
in terms of thermodynamic quantities by introducing a probability density
\begin{align}
 &P(E',h'|E, h) \nonumber\\
\equiv & \frac
    {\sum_{m,n}  |\braket{m,h'}{n,h}|^2 \chi_{I_{E'}}(E(m,h'))\chi_{I_E}(E(n,h)) }
    {\Delta \sum_{n}\chi_{I_E}(E(n,h)) }.
    \label{P-def}
\end{align}
$P(E',h'|E, h) \Delta$ is the probability of finding
the energy in $I_{E'}$ when we instantaneously change
the field from $h$ to $h'$ under the condition that 
the energy eigenstates satisfying $E(n,h) \in I_{E}$ are prepared 
with equal probability. 
From (\ref{qdef new}) and (\ref{P-def}) we have 
\begin{equation}
  \sum_{m,n} |q_{mn} |^2 =  P(E', h'|E,h) D(E,h) \Delta^2,
  \label{q-P}
\end{equation}
from which we can evaluate
$\sum_{m,n} |q_{mn} |^2$ if $P(E',h'|E,h)$ is determined.


Let us fix $P(E',h'|E,h)$. 
By employing the definition (\ref{P-def}) and 
recalling (\ref{m-per}), 
we can find a reasonable form of $P(E',h'|E,h)$ in terms of
\begin{equation}
  \Delta S\equiv S(E',h')-S(E,h). 
\label{DelS}
\end{equation}
We first show in quantum statistical mechanics 
that, for a given $E$ and (\ref{m-per}), 
the most probable transition $E\to E_*'$, 
which maximizes $\log P(E',h'|E,h)$, 
satisfies
\begin{equation}
\Delta S_*=\frac{1}{2}Na(E_{M*},h_M) (\Delta h)^2,
\label{s-star0}
\end{equation}
where $E_{M*}\equiv(E_*'+E)/2$, $h_M\equiv(h'+h)/2$,
and $Na \beta^{-1}$ turns out to be the adiabatic susceptibility \cite{fn:ad}.
See Appendix \ref{quench} for the derivation. 
Then, by using (\ref{s-star0}) and expanding $\log P(E',h'|E,h)$ up to the
second order of $\Delta S$, 
we can express $P(E',h'|E,h)$ as 
\begin{equation}
P(E',h'|E,h) =   e^{-\frac{1}{2 N a (\Delta h)^2}
  \left(\Delta S-\frac{1}{2}N a (\Delta h)^2 \right)^2 +o(N)},
\label{P-final}
\end{equation}
which is derived in Appendix \ref{der-p-final}.

\subsection{Final expression}
By combining (\ref{B_exp3}), (\ref{q-P}) and (\ref{P-final}),
we get the final expression of the propagator
\begin{equation}
B_{j+1}= \Delta^2 e^{-\frac{1}{2 N a_{j+1} (\Delta h_{j+1})^2}
  \left(\Delta S_{j+1}-\frac{1}{2}N a_{j+1} (\Delta h_{j+1})^2 \right)^2 +o(N)}.
\end{equation}
By substituting this into (\ref{mc path bf}),
we obtain the final expression of the path integral in the thermodynamic state space $(E,h)$:
\begin{align}
\nstate{\Psi(t_f)}
=& 
	\left[ \prod_{j=1}^{M} \int_{{\cal D}(h_j)} d E_j
	\right] 
\nstate{\mathcal{P}_{{\bv{E}}_M,{\bv{h}}_M} }{t_f} 
\nonumber \\ 	
&\prod_{j=1}^M 		
	e^{-\frac{i}{\hbar} E_{j-1}\Delta t-\frac{1}{4 Na (\Delta h_j)^2}
        \left(\Delta S_j-\frac{1}{2}N a_j (\Delta h_j)^2\right)^2 +o(N)}.
	\label{mc path final}
\end{align}
We call this a {\it thermodyanamical path integral}. 
This formula can be applied to a class of non-equilibrium processes which are consistent with the two assumptions. 

Apparently, $\nstate{\mathcal{P}_{{\bv{E}}_M,{\bv{h}}_M} }{t_f} $ depends
on  path $(E_1, \cdots, E_{M-1})$. More precisely, (\ref{ETH}) indicates
that $\log |\bnbraket{n,h_M}{\mathcal{P}_{{\bv{E}}_M,{\bv{h}}_M} }{t_f}|^2$ 
is independent of paths when $o(N)$ contribution is ignored, while 
$\psi\equiv {\rm Arg} [\bnbraket{n,h_M}{\mathcal{P}_{{\bv{E}}_M,{\bv{h}}_M} }{t_f}]$
may be path-dependent. With the choise of (\ref{Pj+1}), the phase shift
of $o(N)$ occurs at each time step. We thus assume that the phase $\psi$
is expressed as a function of $(E_1/N, \cdots, E_{M-1}/N)$. 
Then, in the large $N$ limit, 
the dominant contribution of the path
integral (\ref{mc path final}) may be estimated from the saddle
point of 
\begin{equation}
\sum_{j=1}^{M}\left[
-\frac{i}{\hbar} E_{j-1}\Delta t -
\frac{1}{4 Na (\Delta h_j)^2}\left(\Delta S_j-\frac{1}{2}N a_j (\Delta h_j)^2\right)^2\right].
\end{equation}
In the following, we analyze only these terms as the dominant contribution $O(N)$. 

\section{Emergent symmetry in quasi-static operations} 
\label{TEA}
It would be difficult to take a continuum limit of 
the discretized expression of the path integral (\ref{mc path final}). 
In this section, by introducing a variable $\theta$ and considering slow protocols 
referred to as {\it quasi-static operations}, 
we construct a continuum expression. 
Then, we derive an effective action in a thermodynamical phase space 
and find that a symmetry emerges in the path integral, 
which leads to the entropy conservation.
Finally, we discuss characterizations of the variable $\theta$. 

\subsection{Continuous limit in quasi-static operations}
First, we introduce a dimensionless variable $\theta$ through 
\begin{eqnarray}
&& e^{ - \frac{1}{4N a(\Delta h)^2}\left(\Delta S-\frac{1}{2}N a (\Delta h)^2\right)^2} \nonumber \\
&=&
\int d\theta
e^{ - N a (\Delta h)^2 \theta^2- i \theta \left(\Delta S-\frac{1}{2}N a (\Delta h)^2\right)+o(N)}.
\end{eqnarray}
Then, by substituting this into (\ref{mc path final}), 
we obtain
\begin{equation}
\state{\Psi(t_f)}=  \int {\cal D}E \int {\cal D} \theta
\nstate{\mathcal{P}_{{\bv{E}}_M,{\bv{h}}_M} }{t_f} 
e^{{\cal J}+\frac{i}{\hbar} {\cal I}_{\rm eff}}
  \label{path-final}
\end{equation}
with
\begin{eqnarray}
 {\cal J} &\equiv&  
 \sum_{j=1}^M \left[
   - N a_{j}  (\Delta h_j)^2 \theta_j^2 +o(N)\right], 
\label{J-action} \\
  {\cal I}_{\rm eff}&\equiv& \sum_{j=1}^{M}\left[-E_{j-1} \Delta t
   -\hbar \theta_j \left(\Delta S_j-\frac{1}{2}N a_j (\Delta h_j)^2\right) \right], \nonumber \\
\label{eff-act}
\end{eqnarray}
where $ \int {\cal D}E \int {\cal D} \theta =
\prod_{j=1}^M \int dE_j \int d\theta_j$.
${\cal J}$ determines the amplitude 
and ${\cal I}_{\rm eff}$ is the effective action for $(E(t),\theta(t))$. 

Next, we define quasi-static operations. 
We consider $1\ll M \ll \sqrt{N}$ with $M \Delta t=t_f $ fixed.
For simplicity, we assume that $h_j$ increases monotonically,
i.e., $\Delta h_j/h_j= O(1/M)$. Then, from (\ref{m-per}),
$1/\sqrt{N} \ll \Delta h_j/h_j \ll 1$  holds.
For this $h(t)$, we attempt to construct the quasi-static
operation $h^\epsilon(t)$ such that $h^{\epsilon}(t)= h(\epsilon t )$
is satisfied for  $0 \le t \le t^\epsilon_f\equiv t_f/\epsilon$, 
where $\epsilon$ is a small dimensionless parameter
that characterizes the slowness of the operation.
We define
the discrete protocol as $h^\epsilon(t)=h^\epsilon_j$ for $t_j \le t \le t_{j+1}$, 
where $0 \le j \le M^\epsilon\equiv M/\epsilon$ 
and
\begin{equation}
h^\epsilon_j\equiv (1-\epsilon j +\lfloor \epsilon j \rfloor )
h_{\lfloor \epsilon j \rfloor } + (\epsilon j -\lfloor \epsilon j \rfloor )h_{\lfloor \epsilon j \rfloor +1}.
\end{equation}
Here, $\lfloor x \rfloor$ represents the largest integer less than or equal
to $x\in\mathbb{R}$. Indeed, this $h^{\epsilon}(t)$ satisfies 
  $h^\epsilon_{j+1}-h^\epsilon_{j}
  =\epsilon(h_{\lfloor \epsilon j \rfloor +1}-h_{\lfloor \epsilon j \rfloor })$,
which implies that $h^\epsilon_j$ changes slower than $h_j$ by
the factor $\epsilon$, that is, $h^\epsilon(t)=h(\epsilon t)$
in the continuous limit. 
Note that, in order to use the formula (\ref{path-final}), 
the condition $1/\sqrt{N} \ll \Delta h^\epsilon_j/h^\epsilon_j \ll1$
needs to be satisfied; this leads to $1 \ll M^{\epsilon} \ll \sqrt{N}$. 
Because of this condition and $1 \ll M \ll \sqrt{N}$, 
$\epsilon$ should be small but finite so that 
$M/\sqrt{N} \ll \epsilon \ll 1$.

Let us take the continuous limit of the path integral (\ref{path-final}) 
for such monotonically increasing protocol $(h^\epsilon_j)_{j=1}^{M^\epsilon}$. 
Because $\Delta h^\epsilon_j/h_j^\epsilon =O(1/M^\epsilon)$, 
${\cal J}=\sum_{j=1}^{M^\epsilon} N a_{j}  (\Delta h_j^\epsilon)^2 \theta_j^2$
is estimated as $O(N/M^\epsilon)=O(\epsilon N/M)$. 
It becomes smaller as $\epsilon$ is decreased with $N$ and $M$ fixed. 
Therefore, ${\cal J}$ can be neglected for the quasi-static operations. 
Similarly, the third term of (\ref{eff-act}) is negligible. 
Thus, the path integral (\ref{path-final}) becomes 
\begin{equation}
\state{\Psi(t_f^\epsilon)}=  \int {\cal D}E \int {\cal D} \theta
\nstate{\mathcal{P}_{{\bv{E}}_{M^\epsilon},{\bv{h}^\epsilon}_{M^\epsilon}} }{t_f^\epsilon} 
e^{\frac{i}{\hbar} {\cal I}_{\rm eff}}
  \label{path-final-limit}
\end{equation}
with the effective action of $(E(t),\theta(t))$:
\begin{equation}
  {\cal I}_{\rm eff}= 
  \int_0^{t_f^\epsilon}  dt
      \left[ - E(t)-\hbar  \theta(t)
        \der{S( E(t), h^\epsilon(t))}{t}   \right].
\label{eff-act-con}
\end{equation}

\subsection{Emergent symmetry and entropy conservation}
If we transform the integral variable as
\begin{equation}
\theta(t) \to \theta(t)+\eta,
\label{shift}
\end{equation}
where $\eta$ is a small parameter,
(\ref{path-final-limit}) becomes 
\begin{equation}
\state{\Psi(t_f^\epsilon)}= \int {\cal D}E \int {\cal D} \theta
\nstate{\mathcal{P}_{{\bv{E}}_{M^\epsilon},{\bv{h}}^{\epsilon}_{M^\epsilon}} }{t_f^\epsilon} 
e^{\frac{i}{\hbar} {\cal I}_{\rm eff}-i\eta(S_{M^\epsilon}-S_0)}.
\label{path-eta}
\end{equation}
This means that the symmetry for (\ref{shift}) emerges 
in the path integral (\ref{path-final-limit}) for quasi-static operations. 

Let us find the conservation law which is connected to this symmetry 
by the quantum-mechanical Noether theorem. 
We first introduce the entropy operator by 
\begin{align}
\hat S(h) &\equiv  \log D(\hat H(h),h) \nonumber\\
 &= \sum_{n} \log D(E(n,h),h)\state{n,h}\hspace{-2pt}\bstate{n,h}.
\label{S_ope}
\end{align}
Then, we calculate 
\begin{eqnarray}
  & & \hat S(h^{\epsilon}_{M^\epsilon})
\nstate{\mathcal{P}_{{\bv{E}}_{M^\epsilon},{\bv{h}}^{\epsilon}_{M^\epsilon}} }{t_f^\epsilon} \nonumber\\
&=& \sum_{n} \log D(E(n,h^{\epsilon}_{M^\epsilon}),h^{\epsilon}_{M^\epsilon}) \nonumber \\
  & & \times \chi_{I_{E_{M^\epsilon}}}(E(n,h_{M^\epsilon}))d_{nM^\epsilon}({\bv{E}}_{M^\epsilon},{\bv{h}}^{\epsilon}_{M^\epsilon})
  \state{n,h^{\epsilon}_{M^\epsilon}}   ,   \nonumber \\
  & =&
\log D(E_{M^\epsilon},h^{\epsilon}_{M^\epsilon})\sum_{n}\chi_{I_{E_{M^\epsilon}}}(E(n,h_{M^\epsilon}))  \nonumber \\
& & \times 
d_{nM^\epsilon}({\bv{E}}_{M^\epsilon},{\bv{h}}^{\epsilon}_{M^\epsilon})
  \state{n,h^{\epsilon}_{M^\epsilon}}+o(N)    ,   \nonumber \\
&=& 
  S(E_{M^\epsilon},h^{\epsilon}_{M^\epsilon})
  \nstate{\mathcal{P}_{{\bv{E}}_{M^\epsilon},{\bv{h}}^{\epsilon}_{M^\epsilon}} }{t_f^\epsilon}
   +o(N),
\label{S-eigen}
\end{eqnarray}
where we have used (\ref{def_d}) and employed 
$D(E(n,h),h)=D(E,h)+\partial D/\partial E|_E(E(n,h)-E)=D(E,h)(1+\beta(E,h)O(\Delta))$ 
because $E(n,h)\in I_E$ and $\beta \equiv \partial \log D/\partial E$.  
Differentiating (\ref{path-eta}) with respect to $\eta$ and setting $\eta=0$,
we obtain 
\begin{align}
0 &= \int {\cal D}E \int {\cal D} \theta
\nstate{\mathcal{P}_{{\bv{E}}_{M^\epsilon},{\bv{h}}^{\epsilon}_{M^\epsilon}} }{t_f^\epsilon}
(S_{M^{\epsilon}}-S_0) e^{\frac{i}{\hbar} {\cal I}_{\rm eff}} \nonumber \\
&= \int {\cal D}E \int {\cal D} \theta e^{\frac{i}{\hbar} {\cal I}_{\rm eff}} 
\hat S(h^\epsilon_{M^{\epsilon}})
\nstate{\mathcal{P}_{{\bv{E}}_{M^\epsilon},{\bv{h}}^{\epsilon}_{M^\epsilon}} }{t_f^\epsilon}
\nonumber\\
&~~~~~~~~~~~~~~~~~~~~~~~~~~~~~~~~~~-S_0 | \Psi(t_f^\epsilon)\rangle +o(N),
\end{align}
where 
we have used the fact that $S_0$ is independent of the integration, and 
we have employed (\ref{S-eigen}). 
By multiplying this by $\langle \Psi(t_f^\epsilon)|$ and 
noting $\state{\Psi(0)}=\state{n_0,h_0}$, we have 
\begin{equation}
  \bstate{\Psi(t_f^{\epsilon})}
  \hat S(h^{\epsilon}_{M^{\epsilon}})  \state{\Psi(t_f^{\epsilon})}=
  \bstate{\Psi(0)}\hat S(h^{\epsilon}_{0})  \state{\Psi(0)}+o(N).
\label{conserve}
\end{equation}
Thus, the expectation value of the entropy operator is conserved 
for the quasi-static operations. 

\subsection{Characterization of $\theta$}
We study a mathematical structure of the effective
action (\ref{eff-act-con}), which gives characterizations
of the variable $\theta$. 
First, $E$ has one-to-one correspondence with $S$ through
the thermodynamic relation $S=S(E,h)$ for a given $h$.
We can choose $S(t)$ as an independent variable instead of $ E(t)$.
In this representation, (\ref{eff-act-con}) is expressed as 
\begin{equation}
    {\cal I}_{\rm eff}=
    \int_0^{t_f^\epsilon} dt
        \left[ -E(S(t), h^\epsilon(t))
          -\hbar \theta(t) \der{S(t)}{t} \right],
        \label{eff-act-final}
\end{equation}
where $h^\epsilon(t)$ is a given time-dependent parameter.
This can be seen as a canonical-form action 
with Hamiltonian $E(S(t), h^\epsilon(t))$ and 
canonical variables $(S(t),\hbar \theta(t))$. 
Indeed, the symplectic structure $d(\hbar \theta) \wedge dS$ can be obtained 
by taking the exterior derivative of the surface term 
in a general variation of (\ref{eff-act-final}): 
\begin{equation}
\delta {\cal I}_{\rm eff} =\int_0^{t_f^\epsilon} dt 
\left[ \left(-\beta^{-1}+ \der{\hbar \theta}{t}\right)\delta S
-\der{S}{t}\delta(\hbar\theta)\right] 
-\left. \hbar \theta \delta S\right\vert^{t_f^\epsilon}_0.
\end{equation}
We can see that $\delta {\cal I}_{\rm eff} =0$ is equivalent to
the equations
\begin{align}
 \der{\theta(t)}{t}&=\frac{1}{\hbar \beta(S(t),h^\epsilon(t))}, \label{eom1} \\
 \der{S(t)}{t}&=0, \label{eom2}
\end{align}
with the boundary conditions $\hbar \theta=0$ or $\delta S=0$ at
$t=0$ and $t=t_f^\epsilon$. Note here that the energy (or $S$) at
$t=0$ is fixed for the system we are considering. However, we cannot
impose the energy at $t=t_f$, because there is generically no solutions
for this condition. We thus impose $\theta=0$ at $t=t_f$, which is
possible because of the symmetry proprty (\ref{shift}). 
Since the initial energy is fixed, (\ref{eff-act-final}) may be
called the  \textit{microcanonical effective action 
in the thermodynamical phase space} $(S,\hbar \theta)$. 

In the view of (\ref{eff-act-final}), $\hbar \theta$ is the canonically-conjugate variable to the entropy $S$. 
Previously, such a variable was referred to as {\it thermacy} \cite{Dantzig}, 
and effective actions for perfect fluids were constructed
without microscopic derivation
\cite{Brown93,Kambe}. 
Indeed, our action (\ref{eff-act-final}) takes the 
similar form as the previous ones for the spatially
homogeneous cases; however, in these studies, 
the Planck constant does not appear, 
and $(d\hbar \theta/dt) S$ is included instead of $\hbar \theta (dS/dt)$ 
since $\theta$ is fixed at the both boundaries $t=0$ and $t=t_f^\epsilon$
\cite{Brown93}. This effective action describes a different physical
situation from ours (\ref{eff-act-final}). 
Note also that (\ref{eff-act-final}) is derived from quantum mechanics. 

Next, we discuss the concept of {\it thermal time} $\tau$,
a dimensionless quantity  that parameterizes the flow
generated by $-\log \hat \rho$ with a statistical
state $\hat \rho$ \cite{Rovelli93,Connes-Rovelli94,Roveili-Smerlak,Haggard}.
In particular, $\tau$ is determined by 
\begin{equation}
  \frac{d\hat A}{d\tau }=[\hat A,-\log \hat \rho]/i
\end{equation}  
for Heisenberg operators $\hat A$ satisfying 
$  \frac{d\hat A}{dt} =\frac{1}{i\hbar}[\hat A, \hat H]$. 
When $\hat \rho=e^{-\beta \hat H}/Z$,
\begin{equation}
  \frac{d\hat A}{d\tau} =\hbar\beta \frac{d\hat A}{dt}
\end{equation}
holds. 
Comparing this and (\ref{eom1}) implies that $\theta$ corresponds to the thermal time. 

Finally, expressing (\ref{eom1}) as
$dt = \hbar \beta d\theta$, 
we find that the symmetry for $\theta \to \theta +\eta$, (\ref{shift}), 
is equivalent to that
for 
\begin{equation}
t\to t+ \eta \hbar \beta,
\end{equation}
which appears in a different analysis of classical systems \cite{Sasa-Yokokura}. 
\section{Concluding remarks}
Before ending this paper,
we present a few remarks. 
First, as a working hypothesis for obtaining the thermodynamical path
integral (\ref{mc path final}), 
we made the two assumptions in Sec.~\ref{evaluation b}.
The validity of these needs to be checked by applying them 
to various specific models and studying their properties more. 
We postpone this task. 

Second, in order to evaluate physical quantities,
we have to perform the integration of
$E_j$ and $\theta_j$ in (\ref{path-final}).
Here, considering that each term of ${\cal J}$ and ${\cal I}_{\rm eff}$ is $O(N)$,
one may employ a saddle point method with the analytic continuation
of ${\cal J}+i {\cal I}_{\rm eff}/\hbar$
for complex variables $E_j$ and $\theta_j$.
One can then estimate the integral (\ref{path-final}) for specific models 
and directly confirm the symmetry.
Furthermore, it is an important future problem to study how entropy
is not conserved for fast protocols through the
saddle point estimation of ${\cal J}+i {\cal I}_{\rm eff}/\hbar$. 


Third, we remark on the quantum adiabatic theorem: the amplitude in each energy level remains constant 
(and thus $S$ is kept constant) 
if the operation speed is sufficiently slow 
\cite{Tolman,Kato}.
For excited states of nonintegrable many-body systems, 
such a speed becomes extraordinarily slow, which is $e^{-O(N)}$. 
The reason is as follows. 
Since the number of states in an energy shell is $e^{O(N)}$, the distances between neighboring energy levels are $e^{-O(N)}$. 
The operation time of the quantum adiabatic theorem in many-body system is much slower than the lifetime of the universe. 
Therefore, the quantum adiabatic theorem is insufficient to prove the second law of thermodynamics. 
In our theory, by contrast, the operation speed is so fast
that transitions between different energy levels occur.
Nevertheless, the entropy is conserved in (\ref{conserve})
under such operations. 
It is a natural question how to unify the two theories. 


Finally, we hope that experiments will be conducted
to verify our theory. 
In particular, if one observes an entropic effect
of the effective action, the measurement result is quite
interesting. 
One of the most promising ways is an interference experiment. 
In the path-integral formulation, the action in the path integral represents the phase along the corresponding path. 
We thus expect that paths with different entropies (with the same energy) 
exhibit the interference pattern, which is predictable by our theory. 
In the future, we will propose a design of experiments
for this observation. 

\section*{Acknowledgement}
The authors thank M. Hongo, M. Hotta, N. Shiraishi, and H. Tasaki for
their useful comments.
The present study was supported by
KAKENHI (Nos. 25103002, 18K13550, 17H01148, 15J11250,
and 16H07445) and the RIKEN iTHEMS project.


\appendix

\section{Proof of (\ref{log B})}
\label{PSA}
We  prove (\ref{log B}) for any $(n,j)$.
First, the precise statement of (\ref{log B}) is expressed as a
probability:
\begin{equation}
  \lim_{N \to \infty} {\rm Prob}(|\log |C|^2- \log \overline{|C|^{2}}|/N
  \ge \epsilon ) =0
  \label{LL}
\end{equation}
for any $\epsilon >0$, 
where we 
express $C_{nj}$ as $C$. 
To prove this, for $X\equiv \log |C|^2$, we first show that
\begin{eqnarray}
\overline{X} &=& \log \overline{|C|^{2}}+o(N), \label{r-1}\\
{\overline{X^2}-\overline{X}^2}&=& o(N^2), \label{r-2}
\end{eqnarray}
and then use Chebyshev's inequality.

The strategy to show (\ref{r-1}) and (\ref{r-2}) is to use
\begin{equation}
  \overline{X^l}
  = \left. \frac{\partial^l \overline{|C|^{2K}}}{\partial K^l} \right |_{K=0}.
\label{phiK}
\end{equation}
This is obtained by recalling $d^l(a^x)/dx^l =a^x (\log a)^l$, 
setting $a=|C|^2$ and $x=K$, 
and taking the expectation value.   
Let us estimate $\overline{|C|^{2K}}$. 
For $K=2$, by direct calculation, we confirm
\begin{equation}
  \overline{|C|^{4}} = 
 2 \left [\overline{ \left|C \right|^2 } \right]^2 .
\label{K=2}
\end{equation}
For general $K$, we have
\begin{equation}
  \overline{|C|^{2K}} = 
 K! \left [\overline{ \left|C \right|^2 } \right]^K .
\label{K}
\end{equation}
Thus, noting $\log \overline{ \left|C \right|^2 } =O(N)$, we have
\begin{equation}
  \overline{X} 
 =\left. \frac{\partial \overline{|C|^{2K}}}{\partial K} \right |_{K=0}
 = \log \overline{ \left|C \right|^2 }   +o(N), 
 \label{phiK-1}
\end{equation}
and
\begin{equation}
 \overline{X^2}
=  \left. \frac{\partial^2 \overline{|C|^{2K}}}{\partial K^2} \right |_{K=0}
= \left(\log \overline{ \left|C \right|^2 } \right)^2  +o(N^2),
 \label{phiK-2}
\end{equation}
for large $N$.  
From these results, we obtain (\ref{r-1}) and (\ref{r-2}).
We thus conclude (\ref{LL}).

\section{Derivation of (\ref{s-star0})} \label{quench}
We start with the entropy defined by $S(E,h)\equiv \log D(E,h)$ and
consider the most probable value of the entropy change (\ref{DelS}), $\Delta S_*$, 
for a small parameter change $h \to h+\Delta h$ with preparing an equilibrium state initially. 
We then show in the framework of
quantum statistical mechanics that $\Delta S_*$  is given by (\ref{s-star0}):  
\begin{equation}
  \Delta S_* =\frac{1}{2}Na(\Delta h)^2,
  \label{th-ds}
\end{equation}
where $a$ is a non-negative intensive quantity.
In particular, when the Hamiltonian is a linear function of $h$, 
$a$ is expressed in terms of the adiabatic susceptibility.
This corresponds to (\ref{s-star}). 

We begin with the setup. 
For any operator $\hat A$, we define
the expectation value with respect to the microcanonical ensemble by 
\begin{equation}
  \bra   \hat A \ket^{\rm mc}_{E,h}
  \equiv \frac{\sum_{n} \chi_{I_E}(E(n,h)) \bstate{n,h} \hat A \state{n,h} }
  {\sum_{n} \chi_{I_E}(E(n,h))}.
  \label{mc-def}
\end{equation}
Let us define
\begin{equation}
  \hat M(h)\equiv -\frac{\partial \hat H(h)}{\partial h},
\end{equation}
and set 
\begin{equation}
M(E,h)= \bra   \hat M(h) \ket^{\rm mc}_{E,h}.
\end{equation}
This means that 
we consider $\bra \hat M(h) \ket^{\rm mc}_{E,h}$ 
as the thermodynamic value of magnetization $M(E,h)$. 
Indeed, we can show in quantum statistical mechanics that
\begin{equation}
  \pderf{S}{h}{E}=\beta M,
  \label{t-form}
\end{equation}
which, together with the definition of $\beta$, leads to
\begin{equation}
dS=\beta dE +\beta M dh.
\label{f-law}
\end{equation}
The proof of (\ref{t-form})
is given in the argument below (\ref{omega-def}).

The most probable value of the energy change $\Delta E_*$
for the small parameter change $h \to h+\Delta h$
with preparing an equilibrium state initially is given by
the expectation value of $\hat H(h+\Delta h)-\hat H(h)$ 
with respect to the initial equilibrium state: 
\begin{eqnarray}
  & & \Delta E_*  \nonumber \\
& =&    \bra   \hat H(h+\Delta h) \ket^{\rm mc}_{E,h} -
                 \bra   \hat H(h) \ket^{\rm mc}_{E,h}   \nonumber \\
                 &=& -M (\Delta h) +\frac{1}{2}
                 \bra \pdert{\hat H}{h}\ket^{\rm mc}_{E,h}(\Delta h)^2+
                 O( (\Delta h)^3 ).
\label{small-ad}  
\end{eqnarray}  
For this $\Delta E_*$, 
we consider the entropy change $\Delta S_*\equiv S(E+\Delta E_*, h+\Delta h)-S(E,h)$ 
and expand it in $\Delta E_*$ and $\Delta h$. Then, using (\ref{small-ad}), we have 
\begin{eqnarray}
& &  \Delta S_*  \nonumber \\
&=&
 \pderf{S}{E}{h} \Delta E_*+  \pderf{S}{h}{E} \Delta h  \nonumber \\
& & +  \frac{1}{2}
 \left[ \pdertf{S}{E}{h} (\Delta E_*)^2
   + \pdertf{S}{h}{E} (\Delta h)^2 \right]\nonumber \\
& & + \pderc{S}{E}{h} (\Delta E_*)(\Delta h)+O((\Delta h)^3) \nonumber \\
&=& \frac{1}{2}Na (\Delta h)^2+O((\Delta h)^3),
\label{expand}
\end{eqnarray}
where 
\begin{eqnarray}
 N a &\equiv &  M^2\pdertf{S}{E}{h} 
     -2M \pderc{S}{E}{h}
     + \pdertf{S}{h}{E} \nonumber \\
 &+& \beta\bra \pdert{\hat H}{h}\ket^{\rm mc}_{E,h}.
     \label{1st}
\end{eqnarray}

From now, we express $a$ in terms of experimentally measurable quantities.
We start with the identity
\begin{eqnarray}
  &&  \beta \pderf{M}{h}{S} \nonumber \\
  &=& \beta \left|\pder{(M,S)}{(h,E)}\right|
   \left|\pder{(h,E)}{(h,S)}\right|
       \nonumber \\
       &=& \pderf{M}{h}{E}\pderf{S}{E}{h}-\pderf{S}{h}{E}\pderf{M}{E}{h}.
       \label{ident}
\end{eqnarray} 
Here, we notice 
\begin{eqnarray}
  \beta \pderf{M}{h}{E}&=&  \pderf{\beta M}{h}{E}-M\pderf{\beta}{h}{E}
      \nonumber \\
      &=&  \pdertf{S}{h}{E}-M\pderc{S}{h}{E},
      \label{note-1}
\end{eqnarray}
and 
\begin{eqnarray}
  \beta \pderf{M}{E}{h}&=&  \pderf{\beta M}{E}{h}-M\pderf{\beta}{E}{h}
      \nonumber \\
      &=&  \pderc{S}{E}{h}-M\pdertf{S}{E}{h},
      \label{note-2}
\end{eqnarray}
where we have used (\ref{f-law}). 
We substitute (\ref{note-1}) and (\ref{note-2}) into (\ref{ident}),
compare the result with  (\ref{1st}), and then find 
\begin{equation}
  Na=  \beta \pderf{ M}{h}{S}-\beta\bra \pder{\hat M}{h}\ket^{\rm mc}_{E,h}.
\label{Na}
\end{equation}
Note that $a \ge 0$ holds because 
\begin{eqnarray}
  && Na \nonumber \\
&\ge&   \beta \frac{\sum_{n} \chi_{I_E}(E(n,h))
    |\bstate{n,h} (\hat M(h)-M) \state{n,h}|^2 }{ D(E,h)\Delta} \nonumber \\
 & \ge&  0,
\label{Na-2}
\end{eqnarray}
which is shown in Sec. \ref{proof-Na2}.

Finally, we consider the case where 
the Hamiltonian is a linear function of $h$
(as studied in many examples in statistical mechanics). 
Then, since the second term in the right-hand side of (\ref{Na}) vanishes, 
$N a \beta^{-1} $ is the adiabatic susceptibility:
\begin{equation}
Na\beta^{-1}= \pderf{ M}{h}{S}=-\pdertf{E}{h}{S},
\end{equation}
where we have used (\ref{f-law}) at the last equality. 
Following a standard assumption for statistical mechanical models,
we assume that Hamiltonians we study lead to the concavity of $E(S,h)$ in $h$, 
and then we conclude again that $a \ge 0$.

\subsection{proof of (\ref{t-form})}
Let $\Omega(E,h)$ be the number of eigenstates whose
eigenvalues are less than $E$ for Hamiltonian $\hat H(h)$.
That is, 
\begin{equation}
\Omega(E,h) \equiv \sum_{n} \chi(E > E(n,h)),
\label{omega-def}
\end{equation}
where $\chi( X )=1$ if $X$ holds and $\chi( X )=0$ otherwise.
From this definition, we obtain
\begin{eqnarray}
&& \frac{\Omega(E,h+\Delta h)-  \Omega(E,h)}{\Omega(E,h)} \nonumber \\
  & =& \frac{1}{\Omega}\left[
      \sum_{n} \chi(  E(n,h+\Delta h) <E < E(n,h)) \right. \nonumber \\
 & & \left.  -\sum_{n} \chi(  E(n,h) <E < E(n,h+\Delta h)) \right].
\label{start-ent}
\end{eqnarray}
For a small $\Delta h$, the right-hand side can be evaluated as 
\begin{equation}
  - \frac{D(E,h)}{\Omega(E,h)}
  \left. \pder{E(n,h)}{h} \right\vert_{E(n,h)\in I_E}  (\Delta h )
  +o(N),
\label{step}
\end{equation}
where is independent of $n$ satisfying $E(n,h)\in I_E$. 
Since the typical value of 
$\partial E(n,h)/\partial h$ in the energy shell $I_E$
may be replaced by the expectation value with respect to
the microcanonical ensemble,  we have 
\begin{equation}
   \left. \pder{E(n,h)}{h} \right\vert_{E(n,h)\in I_E}
   =  \frac{\sum_{n} \pder{E(n,h)}{h}\chi_{I_E}(E(n,h))}
        {\sum_{n} \chi_{I_E}(E(n,h))}+o(N).
\label{stat}
\end{equation}
By combining this with the identity
\begin{equation}
\bstate{n,h} \frac{\partial \hat H(h)}{\partial h} \state{n,h}  
= \pder{E(n,h)}{h},
\end{equation}
we obtain
\begin{equation}
\left. \pder{E(n,h)}{h} \right\vert_{E(n,h)\in I_E}
=  -M(E,h) +o(N).
\end{equation}
Thus, (\ref{step}) becomes
\begin{equation}
\frac{D(E,h)}{\Omega(E,h)}M (E,h) (\Delta h ) +o(N).
\label{step2}
\end{equation}
By recalling $S(E,h)\equiv \log D(E,h)
=\log \Omega(E,h)+o(N)$ and $\beta(E,h)= D(E,h)/\Omega(E,h)$,
we can re-express (\ref{start-ent}) as
\begin{equation}
  \pder{S(E,h)}{h}=\beta (E,h)M (E,h) 
\end{equation}
which is (\ref{t-form}).


\subsection{proof of (\ref{Na-2})}\label{proof-Na2}

We fix $(E,h)$. For a given small $\Delta h$, we choose $\Delta E$ 
such that $S(E,h)=S(E+\Delta E, h+\Delta h)$. This means
\begin{equation}
  \Delta E+M \Delta h=O( (\Delta h)^2).
\label{ad-ene}  
\end{equation}
For this $\Delta E$, we can have 
\begin{eqnarray}
 & &  \pderf{ M}{h}{S} \Delta h \nonumber \\
& =&  
  \bra \hat M(h+\Delta h)  \ket_{E+\Delta E, h+\Delta h}^{\rm mc}
  -  \bra \hat M(h)  \ket_{E, h}^{\rm mc} \nonumber \\
& &   +O( (\Delta h)^2) .
\end{eqnarray}
From this and (\ref{Na}), we have 
\begin{eqnarray}
&&   Na\beta^{-1} \Delta h \nonumber \\
&=& 
\bra \hat M(h+\Delta h)  \ket_{E+\Delta E, h+\Delta h}^{\rm mc}
- \bra \hat M(h)  \ket_{E, h}^{\rm mc} \nonumber \\
& & 
-\bra \pder{\hat M}{h}\ket^{\rm mc}_{E,h} \Delta h +O( (\Delta h)^2),
\nonumber \\
&=& 
  \bra \hat M(h)  \ket_{E+\Delta E, h+\Delta h}^{\rm mc}
- \bra \hat M(h)  \ket_{E, h}^{\rm mc}+O( (\Delta h)^2).
\label{Na-3}
\end{eqnarray}

Now, we recall (\ref{mc-def}) and re-express it as
\begin{equation}
  \bra   \hat A \ket^{\rm mc}_{E,h}
  =\frac{\sum_{n} \chi(E(n,h) < E) \bstate{n,h} \hat A \state{n,h} }
  {\Omega(E,h)}+o(N)
  \label{mc-def-2}
\end{equation}
for any extensive variable $\hat A$. We start with
\begin{eqnarray}
& & \bra   \hat M(h) \ket^{\rm mc}_{E+\Delta E,h+h+\Delta h} \nonumber \\
  &=& \sum_{n} \frac{\chi(E(n,h+\Delta h) < E+\Delta E)}
                    {\Omega(E+\Delta E,h+\Delta h)} \nonumber \\
  & &  \times  \bstate{n,h+\Delta h} \hat M(h) \state{n,h+\Delta h} 
      +o(N).
  \label{start-M}
\end{eqnarray}
We then have
\begin{eqnarray}
& &  \bra   \hat M(h) \ket^{\rm mc}_{E+\Delta E,h+h+\Delta h}-
  \bra   \hat M(h) \ket^{\rm mc}_{E,h} \nonumber \\
&=& (\Delta M)_1+  (\Delta M)_2+(\Delta M)_3+o(N),
\label{DDDM}
\end{eqnarray}
where
\begin{eqnarray}
& &    \Omega(E,h)  (\Delta M)_1 \nonumber \\
&=&    \sum_{n}
     \left[ \chi(E(n,h) < E+\Delta E-(\pder{E(n,h)}{h} \Delta h) \right.
                                 \nonumber \\
   & &   \left.   -\chi(E(n,h) < E) \right]
     \bstate{n,h} \hat M(h) \state{n,h} , 
 \label{decomp-1} \\
& & \Omega(E,h)  (\Delta M)_2 \nonumber \\
&=& \sum_{n} \chi(E(n,h) < E) 
 \left [\bstate{n,h+\Delta h} \hat M(h) \state{n,h+\Delta h}
            \right. \nonumber \\
& &  \left.    -\bstate{n,h} \hat M(h) \state{n,h} \right]  ,
     \label{decomp-2} \\
& & \Omega(E,h)(\Delta M)_3 \nonumber \\
&=& -\sum_{n} \frac{\chi(E(n,h) < E)}{\Omega(E,h)}
   \nonumber \\
 & & \times \bstate{n,h} \hat M(h) \state{n,h}
              \pder{\Omega}{E}\Delta E+\pder{\Omega}{h} \Delta h.
\label{decomp-3}
\end{eqnarray}
We first see 
\begin{equation}
  (\Delta M)_3 =  -\beta M\Delta E-\beta M^2 \Delta h,
\end{equation}
and find that $(\Delta M)_3 =0$ for $\Delta E$
satisfying (\ref{ad-ene}). 
We then calculate $(\Delta M)_1 $ as 
\begin{eqnarray}
  && (\Delta M)_1  \nonumber \\
  &=& \frac{1}{\Delta}
  \sum_{n} \frac{\chi_{I_E}(E(n,h))}{\Omega(E,h)} \nonumber \\
 &&  \times   \left[\Delta E-\pder{E(n,h)}{h} \Delta h \right] 
    \bstate{n,h} \hat M(h) \state{n,h} \nonumber \\
  &=& \frac{D(E,h)}{\Omega(E,h)}
      \left[M \Delta E \phantom{\frac{1}{D}}  \right. \nonumber \\
 & &      \left.  + \frac{\sum_{n} \chi_{I_E}(E(n,h))
      (\bstate{n,h} \hat M(h) \state{n,h})^2 }{ D(E,h)\Delta}\Delta h \right]
  \nonumber \\
  &=& \beta \Delta h
  \left[-M^2  \phantom{\frac{1}{D}} \right. \nonumber \\
&&     + \left. \frac{\sum_{n} \chi_{I_E}(E(n,h))
      (\bstate{n,h} \hat M(h) \state{n,h})^2 }{ D(E,h)\Delta}\right],
  \label{M1}  
\end{eqnarray}
where we have used (\ref{ad-ene}). 
Next, in order to evaluate $(\Delta  M)_2$, we consider
\begin{eqnarray}
&&  \bstate{n,h+\Delta h} \hat M(h) \state{n,h+\Delta h}
 - \bstate{n,h} \hat M(h) \state{n,h} \nonumber \\
 & =&  \bstate{n,h} \hat M(h)\frac{d}{dh} \state{n,h}\Delta h+({\rm c.c.})
  \nonumber \\
& &   +O((\Delta h)^2).
\end{eqnarray}
Noting that
\begin{eqnarray}
  & &  \bstate{m,h} \frac{d}{dh} \state{n,h} \nonumber \\
&=& 
  \left\{
    \begin{array}{l}
      \displaystyle{\frac{1}{E(m,h)-E(n,h)}} \bstate{m,h} \hat M(h)
      \state{n,h}       \\
       \quad {\rm for} \quad m \not =n, \\
             {\rm i} \theta(n,h) \\
        \quad {\rm for} \quad m =n,
    \end{array}
    \right.
\end{eqnarray}
where $\theta(n,h)$ is a real number, we have
\begin{eqnarray}
&& \bstate{n,h} \hat M(h)\frac{d}{dh} \state{n,h} \nonumber \\
&=& \sum_{m;m\neq n} 
\frac{|\bstate{n,h} \hat M \state{m,h}|^2}{E(m,h)-E(n,h)} \nonumber \\
& & - {\rm i} \pder{E(n,h)}{h}\theta(n,h).
\end{eqnarray}
We thus  obtain
\begin{eqnarray}
&& \Omega(E,h)(\Delta M)_2 \nonumber \\
&=& 
2\sum_{nm;n\neq m} \chi(E(n,h) < E)
\frac{|\bstate{m,h} \hat M(h)  \state{n,h}|^2}{E(m,h)-E(n,h)}\Delta h
\nonumber \\
&=& 
2\sum_{nm;n\neq m} \chi(E(n,h) < E)\chi(E(m,h) > E) \nonumber \\
& &  \times 
\frac{|\bstate{m,h} \hat M(h)  \state{n,h}|^2}{E(m,h)-E(n,h)}\Delta h,
\label{M2}
\end{eqnarray}
where the contribution  $\sum_m \chi(E(m,h) < E)\cdots$ vanishes
from the symmetry for the the exchange of $n$ and $m$. 
Thus, from (\ref{M1}) and (\ref{M2}), we express (\ref{DDDM}) as
\begin{eqnarray}
&&   \bra   \hat M(h) \ket^{\rm mc}_{E+\Delta E,h+h+\Delta h}-
  \bra   \hat M(h) \ket^{\rm mc}_{E,h} \nonumber \\
& =&  \beta \Delta h
 \left[\frac{\sum_{n} \chi_{I_E}(E(n,h))
   (\bstate{n,h} \hat M(h) \state{n,h})^2 }{ D(E,h)\Delta} \right. \nonumber \\
& &   \left. \phantom{\frac{1}{D}}  -M^2 \right] \nonumber \\
& &   +(\Delta M)_2. 
\end{eqnarray}
By recalling (\ref{Na-3}), we arrive at
\begin{eqnarray}
  & &  Na \nonumber \\
  &=&  \beta 
  \left[\frac{\sum_{n} \chi_{I_E}(E(n,h))
      (\bstate{n,h} \hat M(h) \state{n,h})^2 }{ D(E,h)\Delta}-M^2\right]
          \nonumber \\
& &  +\frac{(\Delta M)_2}{\Delta h}, \nonumber \\
& \ge &
 \beta 
  \left[\frac{\sum_{n} \chi_{I_E}(E(n,h))
      (\bstate{n,h} \hat M(h) \state{n,h})^2 }{ D(E,h)\Delta}-M^2\right],
\end{eqnarray}
because  $(\Delta M)_2/(\Delta h) \ge 0$. 
This leads to (\ref{Na-2}).

\section{Derivation of (\ref{P-final})} \label{der-p-final}

We first decompose $\log P(E', h'|E,h)$ into 
\begin{equation}
\log P(E', h'|E,h)=\phi_{\rm S}(E',h'|E,h)+\phi_{\rm A} (E',h'|E,h)
\label{logP}
\end{equation}
with
\begin{eqnarray}
\phi_{\rm S}(E',h'|E,h) &=& \phi_{\rm S}(E,h|E',h'), \label{phis-s} \\
\phi_{\rm A}(E',h'|E,h) &=& -\phi_{\rm A}(E,h|E',h'). \label{phia-s}
\end{eqnarray}
From the symmetry property
\begin{equation}
P(E', h'|E,h)D(E,h)=P(E, h|E',h')D(E',h'),
\label{sym}
\end{equation}
which can be confirmed directly by the definition (\ref{P-def}),
we can determine 
\begin{equation}
\phi_{\rm A}(E',h'|E,h)=\frac{\Delta S}{2}.
\label{phi_A}
\end{equation}

Next we consider $\phi_{\rm S}(E',h'|E,h)$. 
From (\ref{m-per}) and the physical interpretation 
of (\ref{P-def}), we find that the probability of large
$|E'-E|$ is small. 
Noting that for a given $h$, $E$ has one-to-one correspondence with
$S$ through the thermodynamic relation $S=S(E,h)$
and seeing (\ref{phi_A}), 
we expand $\phi_{\rm S}(E',h'|E,h)$ with respect to $\Delta S$,
instead of $\Delta E\equiv E'-E$. 
Therefore, we ignore contribution of $(\Delta S)^4$ and
higher order terms and write 
\begin{eqnarray}
\phi_{\rm S}(E',h'|E,h)&=& N f_0(\Delta h; E_M,h_M) \nonumber \\
&+ & \frac{1}{N} f_2( \Delta h;E_M,h_M ) (\Delta S)^2+o(N),
\label{phi_S}
\end{eqnarray}
for large $N$. 
Here $f_0$ and $f_2$ are $O(N^0)$ functions of
$\Delta h\equiv h'-h$, $E_M\equiv(E+E')/2$ and $h_M\equiv(h+h')/2$ 
which are even in $\Delta h$. 
The mid-point values 
$E_M$ and $h_M$ have been
introduced so that (\ref{phis-s}) is respected. 

Let's determine $f_0$ and $f_2$. 
We note that $P(E', h'|E,h)$ is the probability that 
in thermally isolated macroscopic systems
an equilibrium state with $E$ becomes one with $E'$ 
by the macroscopic perturbation (\ref{m-per}). 
As mentioned above, the most probable value $E'_*$ for given $E$, $h$
and $h'$ satisfies (\ref{s-star0}). 
Such $E'_*$ is characterized by 
\begin{equation}
\left. \pder{\log P(E', h'|E,h)}{E'}\right|_{E'=E'_*} 
=0.
\end{equation}
Through (\ref{phi_A}) and (\ref{phi_S}), we obtain
\begin{equation}
  \frac{\beta'_*}{2}\left[1+ \frac{4 \Delta S_*}{N} f_2{}_* \right]
    + \frac{N}{2} \left[\left.\pder{f_0}{E_M}\right\vert_*
      +\left. \frac{(\Delta S_*)^2}{N^2} \pder{f_2}{E_M} \right\vert_*\right]
    =  0,
\label{s-star} 
 \end{equation}
where $\beta'=\beta(E',h')$ and $|_*$ represents the evaluation at $E'=E'_*$. 
Here, suppose that $f_0 =O((\Delta h)^{\alpha_0})$ and 
$f_2 =O((\Delta h)^{\alpha_2})$ for small $\Delta h/h$. 
Then, the first, second, third, and fourth term of (\ref{s-star}) have
the $\Delta h$ dependence as $(\Delta h)^0$,  
$(\Delta h)^{2+\alpha_2}$, $(\Delta h)^{\alpha_0}$, and 
$(\Delta h)^{4+\alpha_2}$, respectively. By assuming $\alpha_0 \ge 0$
(otherwise (\ref{phi_S}) would become singular when $\Delta h \to 0$),
we obtain $\alpha_0=2$ and $\alpha_2=-2$.
This leads that each bracket in (\ref{s-star}) vanishes, respectively: 
\begin{align}
f_2{}_*&=-\frac{N}{4(\Delta S_*)}=-\frac{1}{2a_*(\Delta h)^2},  \\
\left.\pder{f_0}{E_M}
\right\vert_* &= - \left.
\frac{(\Delta S_*)^2}{N^2} \pder{f_2}{E_M}\right\vert_* 
= - \left. \frac{1}{8}\pder{a}{E_M}\right\vert_* (\Delta h)^2,
\end{align}
where (\ref{s-star0}) has been used. 
We thus  set 
\begin{eqnarray}
  f_2 ( \Delta h;E_M,h_M )&=&-\frac{1}{2 a (E_M,h_M) (\Delta h)^2}, \\
  f_0 ( \Delta h;E_M,h_M )&=& -\frac{1}{8} a (E_M,h_M) (\Delta h)^2.
\end{eqnarray}
From these and (\ref{logP}), (\ref{phi_A}) and (\ref{phi_S}), 
we obtain (\ref{P-final}).




\begin{thebibliography}{99}
\bibitem{comment_d}
By the definitions (\ref{statePj}) and (\ref{def_d}), 
$d_{mj}$ may depend on $\xi_{mk}$ for $k=1,2,\cdots j-1$ but not on $\xi_{mj}$. 
Therefore, $d_{mj}$ is irrelevant to the average with respect to $\xi_{mj}$, (\ref{ex-rs-exp}).
\bibitem{conserved quantities}
When the Hamiltonian has some conserved quantities, the Hilbert space is separated by the eigenvalues of them. By looking at a sector of the conserved quantities, our theory becomes applicable to this case. 
However, when the number of the conserved quantities is $O(N)$, which is the case of the integrable model, the whole story changes. 
Integrable models are outside of the scope of this paper.
\bibitem{fn:ad}  
    We here assume that the Hamiltonian is a linear function of $h$.
    For more general cases, $a$ is not directly connected to the
    the adiabatic susceptibility, but  we can still prove $a \ge 0$. 
    See Appendix \ref{quench} for  these arguments.
\end{thebibliography}
\end{document}